\documentstyle[11pt]{article}
\begin{document}
\begin{center}
{\Large{\bf Non-Central Potentials and Spherical Harmonics Using 
Supersymmetry and Shape Invariance}}\\
\vspace{1.0 in}
{\large \bf 
Ranabir Dutt$^{a,}${\footnote{rdutt@vbharat.ernet.in}}, 
Asim Gangopadhyaya$^{b,}${\footnote{agangop@orion.it.luc.edu,
asim@uicws.phy.uic.edu}}, 
and 
Uday P. Sukhatme$^{c,}${\footnote{sukhatme@uic.edu}}}
\\
\end{center}

\vspace*{.5in}
\noindent 
a) \hspace*{.2in} {\it Department of Physics,  Visva-Bharati University,\\
   \hspace*{.4in} Santiniketan, West Bengal,  India-731235}\\
\noindent
b) \hspace*{.2in}
{\it Department of Physics, Loyola University Chicago, Chicago, Illinois 60626
}\\
c) \hspace*{.2in}
{\it Department of Physics (m/c 273), University of Illinois at Chicago,\\
\hspace*{.4in}845 W. Taylor Street, Chicago, Illinois 60607-7059}\\
\vspace*{.8in}

\centerline{{\bf Abstract}}
It is shown that the operator methods of supersymmetric quantum mechanics and 
the concept of shape invariance can profitably be used to derive properties of 
spherical harmonics in a simple way. The same operator techniques can also be
applied to several problems with non-central vector and scalar potentials.
As examples, we analyze the bound state spectra of an electron in a Coulomb 
plus an Aharonov-Bohm field and/or in the magnetic field of a Dirac 
monopole.
\newpage

Spherical harmonics are introduced in physics courses while treating the 
Laplacian in spherical polar coordinates. In the context of quantum mechanics, 
the Schr\"odinger equation is separable into radial and angular 
parts if the potential is spherically symmetric. The 
angular piece of the Laplacian operator generates spherical harmonics which 
obey interesting and useful recursive formulae.  With the advent of 
supersymmetric quantum mechanics (SUSYQM)[1-3]
and the idea of shape invariance\cite{Gendenshtein83},  study of potential 
problems in nonrelativistic quantum theory 
has received renewed interest. SUSYQM allows one to determine eigenstates 
of known analytically solvable potentials using algebraic 
operator formalism\cite{Dutt} without ever having to solve the Schr\"odinger 
differential equation by standard series method.  However, the operator method 
has so far been applied only to one dimensional and spherically symmetric 
three dimensional problems.

       The study of exact solutions of the Schr\"odinger equation with a 
vector potential and a non-central scalar potential is of considerable 
interest. In recent years, numerous 
studies[6-10]
have been made in analyzing the bound states of an 
electron in a Coulomb field with simultaneous presence
of Aharonov-Bohm (AB)\cite{Aharonov} field and/or a magnetic Dirac 
monopole\cite{Dirac}.  
In most of these studies,  the eigenvalues 
and eigenfunctions are obtained via separation of variables in spherical or 
other orthogonal curvilinear coordinate systems.  
    The purpose of this paper is to illustrate that the idea of 
supersymmetry and shape invariance can be used to obtain exact solutions of 
such non-central but separable potentials in an algebraic fashion.  In this 
method,  
it emerges that the angular part (as well as the radial part) 
of the Laplacian of the Schr\"odinger 
equation can indeed be dealt with using the idea of shape invariance.
This is a novel method of generating 
interesting recurrence properties of spherical harmonics.   
In standard text books on quantum mechanics\cite{qm-texts},  
properties of spherical harmonics are discussed from a different point of 
view. In this regard our approach is new and instructive in the sense 
that the radial and the angular pieces of the Schr\"odinger equation can both
be treated within the same framework. The present work also gives 
insight into the solvability of certain three-dimensional problems. Basically, 
this is a consequence of both separability of variables 
and the shape invariance of the resulting one-dimensional problems.

\newpage
The Schr\"odinger equation for a particle of charge $e$ in the presence of a 
scalar potential $V(r,\theta,\phi)$ and a vector potential 
$\vec{A}(r,\theta,\phi)$ is 
(in units of $\hbar=2m=1$) 
\begin{equation}
[\{-i \vec{\nabla}-e\vec{A}(r,\theta,\phi)\}^2+V(r,\theta,\phi)] \psi 
= E \psi~~.
\end{equation}
We will consider vector potentials of the form 
$\vec{A} = \frac{F(\theta)}{r\sin\theta} \hat{e}_\phi$, and scalar 
potentials 
of the form $V=V_1(r)+\frac{V_2(\theta)}{r^2}$.
The Schr\"odinger equation reads 
\begin{eqnarray}
-\frac{1}{r} \frac{\partial^2}{\partial r^2}(r \psi)
-\frac{1}{r^2\sin \theta} \frac{\partial}{\partial \theta}
\left( \sin\theta \frac{\partial \psi}{\partial \theta}      \right)
-\frac{1}{r^2\sin^2\theta}  \frac{\partial^2\psi}{\partial \phi^2}&&\nonumber\\
+\left( \frac{e^2 F^2(\theta)}{r^2\sin^2\theta} + V_1(r) + 
\frac{V_2(\theta)}{r^2} \right) \psi 
+ \frac{2ie F(\theta)}{r^2\sin^2\theta} \frac{\partial\psi}{\partial \phi}
&=&E \psi.
\label{rP}
\end{eqnarray}
Eq. (\ref{rP}) permits a solution 
via separation of variables, if one writes the wave function in the form 
\begin{equation}
\psi(r,\theta,\phi)=R(r) P(\theta) e^{im \phi}~~.
\end{equation}
The equations satisfied by the quantities $R(r)$ and 
$P(\theta)$ are:

\begin{equation}\label{R}
\frac{d^2R}{dr^2} + \frac{2}{r}\frac{dR}{dr}  +
\left( E-V_1(r)-\frac{l(l+1)}{r^2} \right) R = 0~~, 
\end{equation}
\begin{equation} \label{P}
\frac{d^2P}{d\theta^2} + \cot \theta \frac{dP}{d\theta}  +
\left[l(l+1)- \frac{\left\{ m-F(\theta) \right\}^2}{\sin^2\theta}-V_2(\theta)   
\right]P=0.
\end{equation}
Solutions of these equations for various choices of scalar and vector 
potentials will now be discussed.

\vspace{.2in}
\noindent {\large{\bf Spherical Harmonics:}}
Let us begin with the simplest case of a
free particle $[V_1(r)=V_2(\theta)=F(\theta)=0.]$ This will allow us to 
obtain the standard properties of 
the spherical harmonics. In this case, differential equation (\ref{P}) 
reduces to
\begin{equation}\label{PP}
\frac{d^2P}{d\theta^2} + \cot \theta \frac{dP}{d\theta}
+\left[l(l+1)-\frac{m^2}{\sin^2 \theta}\right]P = 0~~,
\end{equation}
which is the equation satisfied by associated Legendre polynomials. To solve it
by the SUSYQM method, we need to re-cast it into a Schr\"odinger-like 
equation. Changing the 
variable $\theta \rightarrow z$ through a mapping function $\theta = f(z)$, one 
obtains
\begin{equation}
\frac{d^2P}{dz^2} + \left[ -\frac{f''}{f'} + f' \cot f \right]\frac{dP}{dz}
+f'^2 \left[ l(l+1)-\frac{m^2}{\sin^2 f} \right]P = 0~.
\end{equation}
Since we want the first derivative to vanish,  we choose
\begin{equation}
\frac{f''}{f'} = f' \cot f~~,
\end{equation}
which gives
\begin{equation}
\theta \equiv f=2 \tan^{-1}(e^z)~~. \label{theta}
\end{equation}
This transformation amounts to the replacement $\sin\theta={\rm sech} z$ 
and $\cos\theta=-\tanh z$.
The range of the variable $z$ is $-\infty <z< \infty$.
Eq. (\ref{PP}) now reads
\begin{equation}\label{Poschl-T}
-\frac{d^2P}{dz^2}-l(l+1) {\rm sech}^2 z~P =- m^2~ P ~~.
\end{equation}
This is a well-known, shape invariant, exactly solvable potential. It can be
readily solved by standard SUSYQM techniques \cite{Cooper95,Dutt}. The energy 
eigenvalues for the potential $V(z)=-a_0(a_0+1) {\rm sech}^2 z~~(a_0>0)$ are:
\begin{equation}
E_n =-(a_0-n)^2~~; ~~~(n=0,1,2,\cdots N)~~,
\end{equation}
where $N$ is the number of bound states this potential holds, and is 
equal to the largest integer contained in $a_0$. The eigenfunctions 
$\psi_n(z,a_0)$ 
are obtained by using supersymmetry operators\cite{Dutt}:
\begin{equation}\label{dagger}
\psi_n(z;a_0) \propto A^{\dagger}(z;a_0) A^{\dagger}(z;a_1) 
\cdots A^{\dagger}(z;a_{n-1}) \psi_0(z;a_n) ~~,
\end{equation}
where $A^{\dagger}(z;a) \equiv \left( -\frac{d}{dz}+a \tanh z\right)$ 
and $a_n=a_0-n$.
The ground state wave function is $\psi_0(z;a_n)= {\rm sech}^{a_n} z$.

For our problem, $a_0=l$, $E_n=-m^2$ and consequently one has
\begin{equation}
n=l-m~;~~~~P_{l,m}(\tanh z) \sim \psi_{l-m}(z;l),
\end{equation}
where $P_{l,m}(\tanh z)$, the solutions of eq. (\ref{PP}),  are the associated 
Legendre polynomials of degree $l$. Now that these $P_{l,m}(z)$ functions 
can be viewed as solutions of a Schr\"odinger equation, we can apply all
the machinery one uses for a quantum mechanical problem. For example, we
know that the parity of the $n$-th eigenfunction of a symmetric potential 
is given by $(-1)^n$, we readily deduce the parity of $P_{l,m}$  to be 
$(-1)^{l-m}=(-1)^{l+m}\cite{Arfken}.$ Also, the application of supersymmetry 
algebra
results in identities that are either not very well known or not easily 
available. With repeated application of the $A^\dagger$ operators, we can 
determine $P_{l,m}$ for a fixed value ($l-m$). 
As an illustration, we explicitly work out all the polynomials for $l-m=2$. 
The lowest polynomial corresponds to $l=2, m=0$. For a general $l$,  
using eq. (\ref{dagger}), one gets
\begin{eqnarray}
P_{l,l-2}(\tanh z) &\sim & \psi_2(z;l) \nonumber \\
&\sim & A^{\dagger}(z;l) A^{\dagger}(z;l-1) \psi_0(z;l-2) \nonumber \\
&\sim & \left( -\frac{d}{dz} +l\tanh z \right)
\left( -\frac{d}{dz} +(l-1) \tanh z \right) {\rm sech}^{l-2} z  \nonumber \\
&\sim & \left[ -1 + (2l-1) \tanh^2 z \right] {\rm sech}^{l-2} z~~.
\end{eqnarray}
A similar procedure is readily applicable for other values of $n$. When one 
converts back to the original variable $\theta~~({\rm sech} z = \sin \theta~;
\tanh z = -\cos \theta)$, the results are:
\begin{eqnarray}
P_{l,l}(\cos \theta) &\sim &\sin^l \theta \\
P_{l,l-1}(\cos \theta) &\sim & \sin^{l-1} \theta \cos \theta\nonumber \\
P_{l,l-2}(\cos \theta) &\sim &
\left[ -1 +(2l-1)\cos^2\theta\right]\sin^{l-2}\theta
\nonumber \\
P_{l,l-3}(\cos \theta) &\sim &
\left[ -3 +(2l-1)\cos^2\theta\right]\sin^{l-3}\theta 
\cos \theta \nonumber 
\\
P_{l,l-4}(\cos \theta) &\sim &\left[ 3 -6(2l-3)\cos^2 \theta + 
(2l-3)(2l-1)\cos^4 \theta\right]\sin^{l-4}\theta \nonumber \\
P_{l,l-5}(\cos \theta) &\sim &\left[ 15 -10 (2l-3) \cos^2\theta + 
(2l-3)(2l-1)\cos^4 \theta
\right]\sin^{l-5}\theta \cos \theta.\nonumber
\end{eqnarray}
These results are not readily available in standard books, although of 
course they can be obtained from the generating function for associated 
Legendre polynomials. Likewise, there are several recurrence relations 
which are easily obtainable via SUSYQM methods. In particular,
by applying $A$ or $A\dagger$  once, we generate 
recurrence relations of varying degrees (differ in $l$):
\begin{eqnarray}
\psi_{l-m}(z,l)=A^\dagger(z,l) \psi_{l-m-1}(z,l-1) &\Longrightarrow&
P_{l,m}(x)=\left((1-x^2)\frac{d}{dx}+l\,x \right) P_{l-1,m} \\
\psi_{l-m}(z,l)=A(z,l-1) \psi_{l-m+1}(z,l+1) &\Longrightarrow&
P_{l,m}(x)=\left((1-x^2)\frac{d}{dx}+(l-1)\,x \right) P_{l+1,m}.\nonumber 
\end{eqnarray}

\vspace{.2in}
\noindent {\large{\bf Non-Zero Vector Potential:}} 
Here we examine the general case with non-zero $F(\theta)$  in eq. (\ref{P}).
In particular, we consider
\begin{equation}
F(\theta)=\frac{{\cal{F}}}{2\pi} + g(1-\cos \theta)~~,
~~ V_1(r)=-\frac{Ze^2}{r}~~,
~~V_2(\theta)=0~~.
\end{equation}
This choice corresponds to the physically interesting problem of the motion of
an electron in a Coulomb field in the presence of an Aharonov-Bohm potential 
$\vec{A}_{AB}=\frac{{\cal{F}}}{{2\pi}r \sin \theta} \hat{e}_\phi$, 
and a Dirac monopole potential 
$\vec{A}_D=\frac{g(1-\cos\theta)}{r \sin \theta} \hat{e}_\phi$\cite{Villalba}.
Both potentials can be chosen to be of arbitrary strength depending 
on the values of the coupling constants ${\cal F}$ and $g$. Again, we 
transform eq. (\ref{P}) via the change of variables from $\theta$ to $z$ given 
in eq. (\ref{theta}). 
The result is
\begin{equation} \label{pnew}
\frac{d^2P}{dz^2} +[(\lambda^2-{\tilde{m}}^2)-V(z)]P=0~~,
\end{equation}
with
\begin{equation}
V(z)=(\lambda^2+q^2)\tanh^2z-2q{\tilde{m}} \tanh z~~,
\end{equation}
and
\begin{equation}
q=-ge~~,~~{\tilde{m}}=\frac{{\cal{F}}e}{2\pi}+ge-m~~,~~\lambda^2=l(l+1)~~.
\end{equation}
The potential $V(z)$ has the form of the Rosen-Morse II potential, 
which is well-known to be shape invariant. More specifically, the potential
\begin{equation}
V_{RM}(z)=a_0(a_0+1) \tanh ^2 z +2b_0\tanh z~~~(b_0<a_0^2)
\end{equation}
has energy eigenvalues and eigenfunctions\cite{Dutt}
\begin{eqnarray} \label{en}
E_n&=&a_0(a_0+1) - (a_0-n)^2 - \frac{b_0^2}{(a_0-n)^2},  \\
\psi_n&=&(1-\tanh z)^{\frac{1}{2}\left[a_0-n+\frac{b_0}{a_0-n}\right]}
(1+\tanh z)^{\frac{1}{2}\left[a_0-n-\frac{b_0}{a_0-n}\right]}
P_n^{\left( a_0-n+\frac{b_0}{a_0-n},
a_0-n-\frac{b_0}{a_0-n}\right)}.\nonumber
\end{eqnarray}
For our case, $E_n=\lambda^2-{\tilde{m}}^2$ and the constants $a_0$, and $b_0$ 
are given by
$$a_0=-\frac{1}{2}+\sqrt{\frac{1}{4}+(\lambda^2+q^2)}~, ~~~~
b_0=-q{\tilde{m}}~~.$$
Using these values and stipulating that eigenfunctions given in eq. (\ref{en}) 
be normalizable, we get 
\begin{eqnarray}
n&=\sqrt{\frac{1}{4}+(\lambda^2+q^2)} ~~~~ - |q|-
\frac{1}{2}&~~~~~~~~(|q|>|{\tilde{m}}|)~~,\nonumber\\
&=\sqrt{\frac{1}{4}+(\lambda^2+q^2)} ~~~~ - |{\tilde{m}}|-
\frac{1}{2}&~~~~~~~~(|q|<|{\tilde{m}}|)~~.
\end{eqnarray}
Corresponding to these two cases, one gets 
\begin{eqnarray}
l & =-\frac{1}{2} + \sqrt{(n+|q|+\frac{1}{2})^2-q^2} 
& ~~~~~~~~(|q|>|{\tilde{m}}|)~~,\nonumber\\
&=-\frac{1}{2} + \sqrt{(n+|{\tilde{m}}|+\frac{1}{2})^2-q^2} 
&~~~~~~~~(|q|<|{\tilde{m}}|)~~.
\end{eqnarray}
The energy eigenvalues obtained from eq. (\ref{R}) for the Coulomb potential are
\begin{equation} \label{Coul}
E_N=\frac{-Z^2e^4}{4[N+l+1]^2}~~.
\end{equation}
Therefore, our final eigenvalues for a bound electron in a 
Coulomb potential as well a combination of Aharonov-Bohm and Dirac monopole 
vector potentials are
\begin{eqnarray}
E_N&=\frac{-Z^2e^4}{4 \left[ N + \frac{1}{2} + 
\sqrt{(n+|q|+\frac{1}{2})^2-q^2} \right]^2} &
~~~~~~~~ (|q|>|{\tilde{m}}|)~~,\nonumber\\
&=\frac{-Z^2e^4}{4 \left[ N + \frac{1}{2}  + 
\sqrt{(n+|{\tilde{m}}|+\frac{1}{2})^2-q^2} \right]^2} 
&~~~~~~~~ (|q|<|{\tilde{m}}|)~~,
\end{eqnarray}
which agree with eqs. (32) and (33) respectively of ref. \cite{Villalba}.

The above calculation using operator techniques in SUSYQM can 
also be carried out when the Coulomb potential is 
replaced by a harmonic oscillator. Potentials of this type have been 
studied by many authors using other approaches\cite{Kibler93}.

\vspace{.2in}
\noindent {\large{\bf Non-Central Scalar Potential:}}
As a final example, we consider the case of zero vector potential and a scalar 
piece consisting of the Coulomb potential and a
non-central part $V_2(\theta)$. The angular piece satisfies a modified 
version of eq. (\ref{Poschl-T}):
\begin{equation}\label{Poschl-T1}
\frac{d^2P}{dz^2}+\left[ {\rm sech}^2 z \left\{ l(l+1)-V_2(z) \right\}
-m^2\right]~P=0 ~~.
\end{equation}
This equation is the Schr\"odinger equation for one of the known 
shape invariant potentials provided the function $V_2(z)$ is chosen
appropriately. The following three simple choices can be made:
$V_2(z) =  b\sinh z , b\sinh 2z , b\cosh^2z.$ These choices give rise to 
the Scarf II and Rosen-Morse II potentials\cite{Cooper95} 
in eq. (\ref{Poschl-T1}).
In terms of $\theta$, these choices correspond to non-central potentials with 
angular dependences
$\cot \theta, \cot \theta {\rm cosec} \theta {\rm ~and~}
{\rm cosec}^2 \theta$, respectively.

Here, we will treat the case $V_2(z)=b \sinh z$ in detail. 
The full potential in eq. (\ref{Poschl-T1}) is
$-l(l+1) {\rm sech}^2 z + b {\rm sech} z \tanh z$ and the role of 
energy is played by $-m^2$. Recall\cite{Cooper95} that the potential
$V(x)=(B^2-A^2-A) {\rm sech}^2 x + B(2A+1) {\rm sech} x \tanh x$ with $A>0$ 
has eigenvalues 
$E_n=2An-n^2~~(n<A)~$. In terms of our parameters, this implies
\begin{equation}
B(2A+1)=b~~,~~b^2-A^2-A=-l(l+1)~~,~~2An-n^2=-m^2~~.
\end{equation}
Elimination of $A,B$ leads to 
\begin{equation}
l=-\frac{1}{2}+\sqrt{\frac{1}{4}+X}
\end{equation}
with
\begin{equation}
X=\frac{(n^2-m^2)}{2n}+\frac{(n^2-m^2)^2}{4n^2}-\frac{n^2b^2}
{(n^2-m^2+n)^2}~~.
\end{equation}
Substitution for $l$ into eq. (\ref{Coul}) gives 
the eigenvalues for this problem. 
Although we have explicit energy eigenvalues and eigenfunctions, it is not 
clear whether the non-central potential we have considered has physical 
significance. 

In conclusion, we mention that in this paper we have 
attempted to explore the effectiveness of the SUSYQM operator
methods to obtain analytic solutions of Schr\"odinger systems in more than one 
dimensions. It is clear that the factorization technique which has so far 
been applied to only one dimensional or spherically symmetric three dimensional 
problems, can be equally useful for non-central separable problems for which
one dimensional equations can be recast into Schr\"odinger equations with 
shape invariant potentials. 
Many interesting properties of spherical harmonics
emerge naturally as a simple realization of this operator technique.
One should also note that although, in this paper,
we have focused on spherical polar coordinates, any orthogonal curvilinear
coordinate system which is separable into Schr\"odinger-type equations with
shape invariant potentials will allow similar algebraic analysis, and will
have analytically solvable eigenvalues.

This work was supported in part by the U.S. Department of Energy. Two of us,
R.D. and A.G.are grateful to the Physics Department of University of Illinois
at Chicago for kind hospitality. The authors are grateful to Dr. Avinash 
Khare for useful conversations.


\newpage
\begin{thebibliography}{99}
\bibitem{Witten} E. Witten, Nucl. Phys. {\bf B188} (1981) 513,
~Nucl. Phys. {\bf B202} (1982) 253.
\bibitem{Cooper} F. Cooper and B. Freedman, Ann. Phys. {\bf 146} (1983) 262.
\bibitem{Cooper95} For a recent review of supersymmetric quantum mechanics 
and additional references see F. Cooper, A. Khare and U. Sukhatme, Phys. 
Rep. {\bf 251} (1995) 267.
\bibitem{Gendenshtein83} L. Gendenshtein, JETP Letters {\bf 38} (1983) 356.
\bibitem{Dutt} R. Dutt, A. Khare and U.P. Sukhatme, Am. Jour. Phys.,
{\bf 56} (1988) 163.
\bibitem{Kibler87} M. Kibler and T. Negadi, Phys. Lett. {\bf A 124} (1987) 42.
\bibitem{Guha} A. Guha and S. Mukherjee, Jour. Math. Phys. {\bf 28} (1989) 840.
\bibitem{Draganescu} Gh.E. Draganescu, C. Campiogotto, and M. Kibler, 
Phys. Lett. {\bf A 170} (1992) 339.
\bibitem{Kibler93} M. Kibler and C. Campiogotto, Phys. Lett. {\bf A 181} 
(1993) 1 and references contained therein.
\bibitem{Villalba} V.M. Villalba, Phys. Lett. {\bf A 193} (1994) 218.
\bibitem{Aharonov} Y. Aharonov, and D. Bohm, Phys. Rev. {\bf 115} (1959) 485.
\bibitem{Dirac} P.A.M. Dirac, Proc. R. Soc. London, {\bf A 133} (1931) 60.
\bibitem{qm-texts} L. I. Schiff, {\it Quantum Mechanics} 
(3rd Edition), McGraw-Hill,
New York (1968); E. Merzbacher, {\it Quantum Mechanics} (2nd Edition) Wiley,
New York (1970); L. Landau and E. M. Lifshitz, {\it Quantum 
Mechanics, Non-relativistic Theory}, (2nd Edition) Addison-Wesley, 
Reading, Mass. (1965).
\bibitem{Chetonani} L. Chetonani, L. Guechi, and T.F. Harman, Jour. Math. 
Phys. {\bf 30} (1989) 655.
\bibitem{Arfken}Handbook of Mathematical Functions, M. Abramowitz and I.A.
Stegun, Dover Publications, Inc., New York, 1970.\\
Mathematical Methods for Physicists, George Arfken, 
Third Edition, Academic Press.


\end{thebibliography}
\end{document}